\PassOptionsToPackage{usenames, dvipsnames}{xcolor}
\documentclass{article}

\usepackage[final]{neurips_2023}

\usepackage{framed}

\usepackage{listings}
\usepackage{multirow}
\usepackage{array}
\usepackage{orcidlink,thumbpdf,lmodern}
\usepackage[utf8]{inputenc} % allow utf-8 input
\usepackage[T1]{fontenc}    % use 8-bit T1 fonts
\usepackage{hyperref}       % hyperlinks
\usepackage{url}            % simple URL typesetting
\usepackage{booktabs}       % professional-quality tables
\usepackage{amsfonts}       % blackboard math symbols
\usepackage{nicefrac}       % compact symbols for 1/2, etc.
\usepackage{microtype}      % microtypography
\usepackage{xcolor}         % colors
\title{pyKCN: \\ A $Python$ Tool for Bridging Scientific Knowledge}

\author{Zhenyuan Lu \quad Wei Li  \quad Burcu Ozek \quad Haozhou Zhou \AND Srinivasan Radhakrishnan \quad Sagar Kamarthi \vspace{0.3em} \\
{\normalsize \{lu.zhenyua, li.wei10, ozek.b, zhou.haoz, s.kamarthi, s.radhakrishnan\}@northeastern.edu} \quad \\
{\normalsize Northeastern University, Boston} \quad \\
}

\begin{document}

\maketitle

\begin{abstract}
%% General background
The study of research trends is pivotal for understanding scientific development on specific topics. 
%% Specific introduction and gap
Traditionally, this involves keyword analysis within scholarly literature, yet comprehensive tools for such analysis are scarce, especially those capable of parsing large datasets with precision. 
%% Our contribution (fills the gap)
$pyKCN$, a Python toolkit, addresses this gap by automating keyword cleaning, extraction and trend analysis from extensive academic corpora. 
%% Significant outcome
It is equipped with modules for text processing, deduplication, extraction, and advanced keyword co-occurrence and analysis, providing a granular view of research trends. This toolkit stands out by enabling researchers to visualize keyword relationships, thereby identifying seminal works and emerging trends. Its application spans diverse domains, enhancing scholars' capacity to understand developments within their fields. 
%% Implication
The implications of using $pyKCN$ are significant. It offers an empirical basis for predicting research trends, which can inform funding directions, policy-making, and academic curricula. 
%% Conclusions
$pyKCN$ thus represents a leap forward in research trend analysis, offering scholars a powerful tool for navigating the knowledge landscape. The code source and details can be found on: \url{https://github.com/zhenyuanlu/pyKCN}
\end{abstract}

\section{Introduction}

Literature reviews can be broadly classified into research landscape analysis and detailed topical reviews. Research landscape analysis, including scientometric and bibliometric reviews, offers a quantitative, comprehensive overview of a research area. It highlights key trends, major themes, and overall trajectories by analyzing the meta-data of a large number of literature through data-driven methods. On the other hand, detailed topical reviews, like systematic reviews, involve rigorous human screening and focus on in-depth analysis of specific topics. They offer detailed insights, critical evaluations, and in-depth discussions on particular findings or methodologies. Both types of reviews serve unique yet complementary purposes. Research landscape analysis helps researchers understand the larger landscape, identify research trends, and recognize potential gaps or underserved areas. Meanwhile, a detailed topical review provides a deeper understanding of specific issues, theories, or methods, facilitating nuanced discussions and fostering specialized expertise.

Automated tools designed to assist literature reviews are increasingly needed in academic research. VOSviewer \cite{van2010software} and Connected Papers \cite{Ammar2018ConstructionOT} are two notable examples of research landscape analysis tools. VOSviewer utilizes bibliometric data to create bibliometric network visualizations, revealing the connections between researchers, institutions, and specific keywords. Similarly, Connected Papers examines the bibliographic data of a selected paper and generates a graphical map that demonstrates how that paper is related to others in its field. As for detailed topical reviews, tools powered by Large Language Models (LLM), such as PDF.ai \cite{pdfai2023}, come into play. These tools take individual research papers or collections, extract key findings, generate summaries, and answer specific queries about the content. 

Despite the efficiency of specialized automated tools, there are still many challenges and limitations. Many tools tend to operate in isolation, focusing either on a broad overview or on detailed insights without offering a comprehensive integration of both. While they provide visualizations and summaries, they lack insightful metrics to gauge the dynamic evolution of research fields. Additionally, most platforms do not track or visualize the shift in research themes over time. In this context, we introduce pyKCN, a computational tool designed to aggregate publication metadata of a research area and generate comprehensive insights. It aims to answer the following questions:

\begin{quote}
1. Is this research area trending? How can we quantify its growth over time?

2. How can we quantitatively describe the research landscape? Is the landscape expanding, gaining depth, or both? Which specific areas are expanding or gaining depth?

3. What are the emerging and declining topics in this research area? How does the research focus shift over time?

4. Which topics are frequently studied together? How do these associations change over time?
\end{quote}

The pyKCN package implements natural language processing (NLP) modules and downstream task modules like keyword co-occurrence network (KCN) analysis. The user downloads the relevant literature metadata from their preferred database to start the analysis. From there, the package constructs a robust keyword database using pipelines powered by natural language processing toolkits. These pipelines perform essential tasks like tokenization, stemming, lemmatization, and merging synonyms and acronyms, ensuring the analysis is not skewed by linguistic habits.

Following this preprocessing, co-occurrence values are calculated to determine how often two keywords appear together in literature. Essentially, KCN works by viewing each keyword as a node and every co-occurrence of a pair of words as a link. The frequency of these co-occurrences establishes the weight of the link, resulting in a weighted network. The package provides metrics to evaluate the nodes and edges' statistical features, centrality, affinity, and cohesiveness. Beyond KCN, pyKCN offers a range of downstream tasks like association rules generation, research trend visualization, and potential integration with advanced technologies like GPT-4. 

The package is separated into distinct modules that perform specific tasks. The first module is the data preparation module, which extracts keyword data from metadata while ensuring data validity. The second module is the text processing module, which uses NLP tools to process the data. The third module is the matrix generator, which generates the network co-occurrence matrix. The final modules are the downstream models, which calculate network metrics and association rules. The package includes meticulous error handling to ensure smooth operation. For performance and versatility, the package utilizes Python libraries such as pandas, numpy, nltk, networkX, and matplotlib.

\section{Related Work}

This section briefly reviews the existing software for literature reviews and compares them with pyKCN. The KCN methodology, introduced in 2017 \cite{radhakrishnan2017novel}, has been cited in over 200 scientometric review papers. In this section, we summarize various research areas where KCN has been applied, validating the need for an integrated tool like pyKCN. Lastly, we discuss pyKCN's key features and their value to the evolving field of scientometric reviews. 

\begin{table}
    \centering
    \renewcommand{\arraystretch}{1.3} % Adjusts the row height
    \begin{tabular}{|>{\centering\arraybackslash}m{3cm}|>{\centering\arraybackslash}m{3.2cm}|>{\centering\arraybackslash}m{2.5cm}|>{\centering\arraybackslash}m{2.2cm}|>{\centering\arraybackslash}m{2.2cm}|} \hline
         \textbf{Main Function} & \textbf{Software Name} & \textbf{ Analysis Capability} & \textbf{Automated Trend Analysis} & \textbf{Natural Language Processing} \\ \hline
         \multirow{3}{3cm}{LLM-powered literature analysis} & pdf.ai & Single paper & No & Yes \\ \cline{2-5}
         & Semantic Reader & Single paper & No & Yes \\ \cline{2-5}
         & Taskade & Both & No & Yes \\ \hline
         \multirow{3}{3cm}{Literature management and screening} & MAXQDA & Both & No & Yes \\ \cline{2-5}
         & DistillerSR & Both & No & Yes \\ \cline{2-5}
         & Rayyan & Both & No & Yes \\ \hline
         \multirow{3}{3cm}{Research landscape analysis and visualization} & VOSviewer & Meta & No & No \\ \cline{2-5}
         & Connected Papers & Meta & No & No \\ \cline{2-5}
         & CiteSpace & Meta & Yes & No \\ \hline
    \end{tabular}
    \caption{Comparison of Literature Review Tools}
    \label{tab:literature_review_tools}
\end{table}

Table 1 presents a range of literature review tools, each serving different purposes in the review process. Tools like pdf.ai represent a category that, while not exclusive to literature reviews, assists in extracting information from publications, relying on manual screening without analyzing research trends. Semantic Reader, another example, offers real-time AI-powered abstracts of cited work and highlights key parts of publications. Taskade, with its flexibility and collaborative features, can be tailored to support literature reviews, though it also lacks automated trend analysis.

In contrast, tools such as MAXQDA, DistillerSR, and Rayyan fall under literature management and screening, designed for detailed topical reviews. MAXQDA focuses on managing and analyzing qualitative data, while DistillerSR and Rayyan are tailored to assist systematic reviews, especially in the medical field.

Lastly, bibliometric network analysis tools like VOSviewer, Connected Papers, and CiteSpace are closely related to pyKCN's domain. They create networks based on citation relationships. pyKCN, on the other hand, creates networks based on contents from title and keywords, leverages NLP to refine raw data and employs KCN methodology to quantify research trends, complementing the capabilities of tools like CiteSpace.

\begin{figure}[t!]
	\centering
	\includegraphics[width=0.7\linewidth]{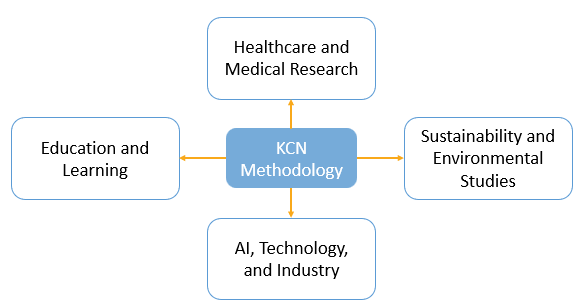}
    \caption{KCN Application}
    \label{fig:kcn_application}
\end{figure}

According to a recent study, the popularity of scientometric reviews is on the rise. Figure 1 summarizes the areas where the KCN has been applied. However, not all such reviews meet the criteria for being valuable and effective. A high-quality scientometric review should ensure input data quality, utilize nuanced search strategies, and include insightful visualization interpretations. It should also conduct comprehensive analyses across multiple dimensions such as intellectual composition and temporal dynamics. pyKCN aligns well with these requirements, offering robust data processing via NLP modules and versatile tools for both structural and temporal analyses. Additionally, its interactive visualizations and advanced metrics aid in identifying key research trends, making pyKCN a valuable asset for conducting impactful scientometric reviews.

\section{Architecture and Core Functionality}

\begin{figure}[t!]
	\centering
	\includegraphics[width=\linewidth]{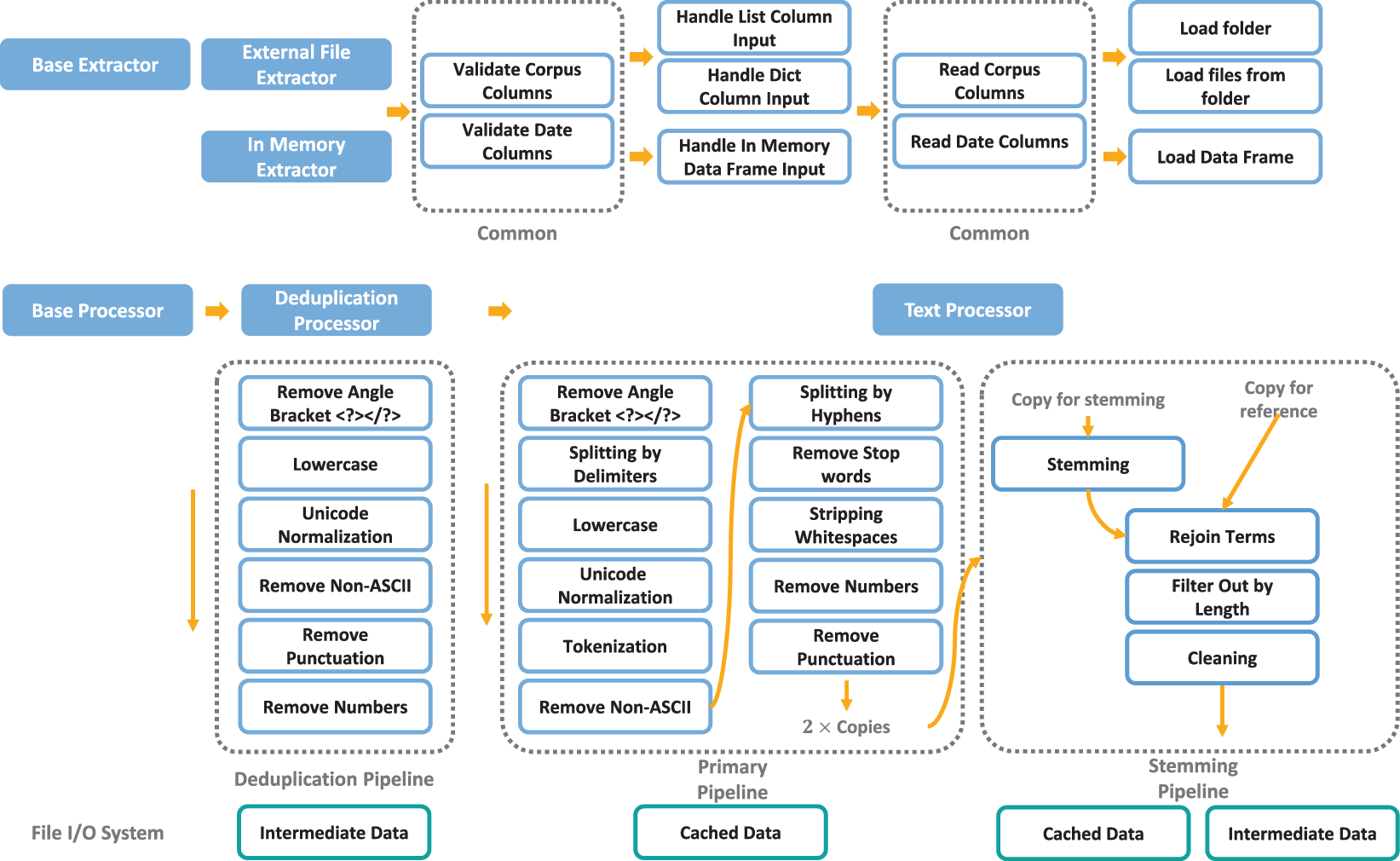}
% 	\vspace{-13pt}
    \caption{The main architecture of pyKCN.}
    \label{fig:pykcn_architecture}
\end{figure}

The \textit{pyKCN} software is architected with modularity and scalability at its core. This section elucidates the overarching blueprint of \textit{pyKCN}, highlighting its key components and their interplay.

\subsection{Overview of the pyKCN Architecture}

At the highest level, \textit{pyKCN}'s architecture can be categorized into primary and auxiliary components \ref{fig:pykcn_architecture}. The primary components serve as the software's main backbone, while the auxiliary components facilitate and enhance the functionalities provided by the primary modules.

\begin{itemize}
    \item \textbf{Main Modules:} These form the core of \textit{pyKCN}, encapsulating the primary functionalities and algorithms. They are designed with flexibility in mind, enabling users to seamlessly integrate and expand upon them as needed.
    
    \item \textbf{Logging System:} An indispensable part of any robust software, the logging system in \textit{pyKCN} ensures that all operations, anomalies, and significant events are meticulously recorded. This aids in debugging, performance tuning, and maintaining transparency in operations.
    
    \item \textbf{File System:} This component is responsible for the structured storage and retrieval of data. By leveraging a hierarchical and organized approach, it ensures efficient data access and modification capabilities.
    
    \item \textbf{Error Handling:} An essential layer that oversees the graceful management of exceptions and unexpected events. It ensures the software's robustness and resilience in the face of unforeseen challenges.
    
    \item \textbf{Docstring, Annotation, and Documentations:} This segment is dedicated to providing comprehensive documentation for \textit{pyKCN}. From method-specific docstrings to overarching annotations, it aids users in understanding and navigating the software effectively.
\end{itemize}

Supplementing these primary components are core modules:

\begin{itemize}
    \item \textbf{Unit/Integration Test:} Safeguarding the integrity of \textit{pyKCN} are rigorous unit and integration tests. These tests ensure that both individual modules and their collective interactions remain flawless and efficient.
    
    \item \textbf{Other Modules:} These encapsulate additional functionalities and tools that can be leveraged to extend the core capabilities of \textit{pyKCN}.
    
    \item \textbf{Library Setup and Configuration:} Facilitating the initial setup and subsequent configurations, this module ensures that users can tailor \textit{pyKCN} to their specific requirements with ease.
    
    \item \textbf{Examples:} A repository of practical use-cases and demonstrations that guide users in harnessing the power of \textit{pyKCN} to its fullest potential.
    
    \item \textbf{Continuous Integration (CI):} Ensuring the seamless integration of updates and modifications, the CI process validates the compatibility and performance of new additions, maintaining the software's integrity.
\end{itemize}

In summation, the \textit{pyKCN} architectural blueprint is a testament to thoughtful design, emphasizing modularity, scalability, and user-centricity. Through its well-defined components and their strategic interplay, \textit{pyKCN} stands poised to address a wide array of tasks and challenges with finesse.

\subsubsection{Main Modules}
The architecture of our main modules can be chiefly decomposed into two pivotal modules: the extraction module and the processing module. Both of these modules synergize to facilitate efficient data management and text processing. Below we introduce the components of these modules, elucidating their roles and interactions.

\textbf{Base Extractor}: This acts as the foundational layer for data extraction. It establishes the general protocol and methods which are then specialized by other extractors.

\begin{itemize}
    \item \textbf{External File Extractor}: Specifically tailored to handle data extraction from external files. This component ensures that files from diverse sources and formats can be seamlessly integrated into our system.
    \item \textbf{In Memory Extractor}: Designed to retrieve data stored in memory. This is particularly vital for instances where data is temporarily cached or buffered.
\end{itemize}

\textbf{Base Processor}: Serves as the backbone for all processing tasks, outlining the fundamental logic and order of operations.

\begin{itemize}
    \item \textbf{Deduplication Processor}: An indispensable unit responsible for identifying and eliminating duplicate data entries. This not only optimizes storage but also ensures the integrity and accuracy of subsequent processing.
    \item \textbf{Text Processor}: This component is entrusted with the crucial task of parsing and processing textual data. It's equipped to make multiple copies for diverse tasks, such as stemming and reference tracking.
\end{itemize}

\textbf{Deduplication Pipeline}: Anchored by the Deduplication Processor, this pipeline rigorously scans the incoming data from the File I/O System to filter out any redundancies.

\textbf{Primary Pipeline}: Post deduplication, the data flows into the primary pipeline. Here, two main copies are generated, setting the stage for specialized processing.

\textbf{Stemming Pipeline}: One of the copies is earmarked for stemming, a linguistic process wherein words are reduced to their base or root form, facilitating uniformity and simplifying subsequent textual analyses.

% ------------------------------
% -- Data Extraction --
% ------------------------------

\subsection{Data Extraction Modules}
\subsubsection{BaseExtractor: A Foundational Framework for Data Extraction}

The \lstinline|BaseExtractor| class is designed as the foundational layer for data extraction, offering robust interfaces and validating mechanisms to ensure a smooth extraction process. The emphasis on a modular architecture becomes evident as it enables easier integration with specialized data extraction approaches and broadens the applicability spectrum.

\begin{lstlisting}[language=Python]
class BaseExtractor:
    """
    Base class for data extraction.
    """
\end{lstlisting}

The constructor of the class, \lstinline|__init__|, accepts various parameters that cater to diverse data extraction scenarios:

\begin{itemize}
    \item \lstinline|data_frame|: Optional data frame where the extracted data will be stored.
    \item \lstinline|data_mapping|: A dictionary detailing the mapping for data extraction.
    \item \lstinline|new_column_names|: A list indicating the new names of columns post-extraction.
    \item \lstinline|data_dir|: A string representing the directory where data resides.
    \item \lstinline|corpus_columns|: A list detailing the columns which need corpus-based processing.
    \item \lstinline|date_column|: Specifies the column that contains date data.
    \item \lstinline|date_type|: Defines the type of date data ('year', 'numeric', or 'string').
\end{itemize}

To ensure a consistent and error-free extraction process, \lstinline|BaseExtractor| incorporates a comprehensive validation mechanism, which can be broken down as follows:

\paragraph{Exclusive Parameter Validation}
This validation ensures that users provide either file input parameters or in-memory DataFrame parameters, but not both simultaneously, eliminating potential conflicts in the extraction process.
\begin{lstlisting}[language=Python]
    def _validate_exclusive_parameters(self):
\end{lstlisting}

\paragraph{Data Mapping and Directory Validation}
This validation is executed if a \lstinline|data_mapping| is provided, ensuring that the accompanying \lstinline|data_dir| is also given.
\begin{lstlisting}[language=Python]
    def _validate_data_mapping_and_dir(self):
\end{lstlisting}

\paragraph{Corpus and Date Column Validation}
It ensures that if a \lstinline|corpus_columns| list is provided, the corresponding \lstinline|date_column| must also be provided.
\begin{lstlisting}[language=Python]
    def _validate_corpus_and_date_columns(self):
\end{lstlisting}

\paragraph{Type and Structure Validation}
It guarantees the correctness of the types of various parameters, ensuring logical coherence and effective data extraction.
\begin{lstlisting}[language=Python]
    def _validate_types(self):
\end{lstlisting}

A salient feature of the \lstinline|BaseExtractor| is its capacity to handle various date formats. The \lstinline|date_extractor| function offers versatile data extraction capabilities, accommodating types like 'year', 'numeric', and 'string', and leveraging different extraction functions based on the specified date type.

The class provides a generic \lstinline|load_data| method that must be overridden by its subclasses to address specific data loading requirements. The pivotal function, \lstinline|extract_data|, is responsible for initiating the extraction process, processing date and corpus columns, and returning the final data frame post-extraction.

In essence, the \lstinline|BaseExtractor| stands as a versatile and robust foundation for data extraction, encapsulating comprehensive validation mechanisms, ensuring robustness, and allowing seamless extensibility.

% ------------------------------
% -- ExternalFileExtractor --
% ------------------------------

\subsubsection{ExternalFileExtractor}

The \lstinline|ExternalFileExtractor| class, extending from the previously discussed \lstinline|BaseExtractor|, offers a robust mechanism to handle data extraction from common file types such as CSV and Excel files. The following subsection delves into the intricacies of this extractor.

\paragraph{External File Extraction}
The process of extracting data from external files, given their ubiquity in datasets, is pivotal. Our solution, the \lstinline|ExternalFileExtractor|, focuses on streamlining this operation, ensuring efficiency and adaptability.

\textit{Constructor and Initial Setup}
The \lstinline|ExternalFileExtractor| is initialized with various parameters to provide flexibility during extraction. Its constructor, \lstinline|__init__|, inherits the parameters from the `BaseExtractor` but also integrates a logger for monitoring operations:
\begin{lstlisting}[language=Python]
class ExternalFileExtractor(BaseExtractor):
    def __init__(self,
                 data_mapping: dict[str] | None = None,
                 data_dir: str = None,
                 new_column_names: list[str] | None = None,
                 date_type: str = 'year'):
        super().__init__(data_mapping = data_mapping,
                         data_dir = data_dir,
                         new_column_names = new_column_names,
                         date_type = date_type)
        self.logger = logging.getLogger(__name__)
\end{lstlisting}

\textit{Data Loading Mechanism}
A critical function in this class, \lstinline|load_data|, orchestrates the entire process of data extraction, offering a consolidated DataFrame from disparate file sources:
\begin{lstlisting}[language=Python]
    def load_data(self) -> tuple[pd.DataFrame, list[str], str]:
        # Code logic...
\end{lstlisting}

It inherently utilizes another method, \lstinline|_load_data_from_folder|, to facilitate folder-specific extraction and handle configuration-based data retrieval:
\begin{lstlisting}[language=Python]
    def _load_data_from_folder(self, folder: str, config: dict) -> tuple[DataFrame | None, list[str], list[str], str]:
        # Code logic...
\end{lstlisting}

\textit{File-Specific Data Extraction}
The extraction mechanism distinguishes between file types, ensuring accurate data retrieval. The method \lstinline|_load_data_from_file| is instrumental in this aspect:
\begin{lstlisting}[language=Python]
    def _load_data_from_file(self,
                             folder_path: str,
                             file: str,
                             target_columns: list[str]) -> pd.DataFrame | None:
        # Code logic...
\end{lstlisting}
This method checks the file extension and decides the appropriate pandas function to deploy, be it \lstinline|read_csv| for CSV files or \lstinline|read_excel| for Excel files.

\textit{Column Renaming and Standardization}
Considering diverse datasets might have varying column names, standardization becomes vital. The \lstinline|_get_new_column_names| and \lstinline|_get_final_column_names| methods cater to this requirement:
\begin{lstlisting}[language=Python]
    def _get_new_column_names(self,
                              corpus_columns: list[str],
                              date_column: str) -> tuple[list[str], list[str], str]:
        # Code logic...

    @staticmethod
    def _get_final_column_names(corpus_columns: list[str],
                                date_column: str,
                                new_column_names: dict) -> tuple[list[str], str]:
        # Code logic...
\end{lstlisting}

\textit{Error Handling and Logging}
Given the intricacies of file handling, potential errors need addressing. The \lstinline|_log_error| method is designed for this very purpose, ensuring that extraction issues are promptly logged:
\begin{lstlisting}[language=Python]
    def _log_error(self, error_message: str) -> None:
        """
        Log an error message.
        :param error_message: The error message to log.
        :return: None
        """
        self.logger.error(error_message)
\end{lstlisting}

\textbf{Summary:}
The \lstinline|ExternalFileExtractor| works as the external file data extraction, emphasizing flexibility, efficiency, and robustness. By seamlessly integrating with the foundational \lstinline|BaseExtractor|, it showcases a pragmatic approach towards diverse dataset handling in the realm of representation learning. Future work might encompass supporting an even broader spectrum of file formats, further augmenting the utility of this extractor in various data science endeavors.

% ------------------------------
% -- InMemoryDataFrameExtractor --
% ------------------------------

\subsubsection{InMemoryDataFrameExtractor}

One increasingly common scenario involves the need to process data that already resides in memory, typically in structured formats like DataFrames. The \lstinline|InMemoryDataFrameExtractor| caters to this very scenario, providing robust extraction capabilities for in-memory data structures.

\paragraph{In-Memory Data Extraction Mechanism}
The \lstinline|InMemoryDataFrameExtractor| builds upon the foundational BaseExtractor class, inheriting its core functionalities and extending them to cater to in-memory data extraction. Its design emphasizes ease of use, flexibility, and error resilience, ensuring that data can be seamlessly loaded and preprocessed, even from complex in-memory structures.

\textit{Initialization}
The class constructor accepts several parameters:
\begin{itemize}
\item \lstinline|data_frame|: The core DataFrame from which data is to be extracted.
\item \lstinline|new_column_names|: Optionally, a list of new column names if renaming is required.
\item \lstinline|corpus_columns|: Columns which constitute the corpus.
\item \lstinline|date_column|: A designated column for date information.
\item \lstinline|date_type|: Specifies the granularity of the date (e.g., 'year').
\end{itemize}

The initialization method instantiates the logger and invokes the superclass's initialization with the provided parameters, as evident from the snippet below:

\begin{lstlisting}[language=Python]
def init(self,
    data_frame: pd.DataFrame,
    new_column_names: list[str] = None,
    corpus_columns: list[str] = None,
    date_column: str = None,
    date_type: str = 'year'):
    super().init(data_frame = data_frame,
    new_column_names = new_column_names,
    corpus_columns = corpus_columns,
    date_column = date_column,
    date_type = date_type, )
    self.logger = logging.getLogger(name)
\end{lstlisting}

\textit{Data Loading and Preprocessing}
The \lstinline|load_data| method is at the heart of the extraction process. It leverages the power of pandas' DataFrame operations to:

Extract relevant columns based on \lstinline|corpus_columns| and \lstinline|date_column|.
Rename the columns if \lstinline|new_column_names| is provided.
Return the final concatenated DataFrame along with the corpus and date columns.
Here's the method in its entirety:

\begin{lstlisting}[language=Python]
def load_data(self) -> tuple[pd.DataFrame, list[str], str]:
try:
    target_columns = self.corpus_columns + [self.date_column]
    final_df = self.data_frame[target_columns]
    final_corpus_columns = self.corpus_columns
    final_date_column = self.date_column

    if self.new_column_names:
        new_column_mapping = dict(zip(target_columns, self.new_column_names))
        final_df.rename(columns = new_column_mapping, inplace = True)
        final_corpus_columns = [new_column_mapping[col] for col in self.corpus_columns]
        final_date_column = new_column_mapping[self.date_column]
        return final_df, final_corpus_columns, final_date_column
    except Exception as e:
        self._log_error(READ_ERROR.format(e))
        return pd.DataFrame(), [], ""
\end{lstlisting}

\textit{Error Handling and Logging}
Given the variable nature of in-memory data structures, there is a potential for extraction errors. The \lstinline|InMemoryDataFrameExtractor| includes a dedicated \lstinline|_log_error| method, which is invoked to log any exceptions encountered during the data extraction process:

\begin{lstlisting}[language=Python]
def _log_error(self, error_message: str) -> None:
    self.logger.error(error_message)
\end{lstlisting}

\textbf{Summary:}
The \lstinline|InMemoryDataFrameExtractor| plays a pivotal role in extracting data from in-memory structures, merging flexibility with robustness. By ensuring seamless integration with pandas DataFrames, this module paves the way for efficient representation learning directly from in-memory data, mitigating the need for external data loading processes.

% ------------------------------
% -- Text Processing --
% ------------------------------
\subsection{Text Processing Modules}

Central to the data processing pipeline of \textit{pyKCN} is the \textbf{BaseProcessor} class. This class encapsulates a plethora of methods meticulously engineered to process, cleanse, and refine data, paving the way for advanced analysis and operations.

\subsubsection{BaseProcessor: A Detailed Exploration}

The \lstinline|BaseProcessor| is initialized with a series of parameters that dictate its behavior across various operations. A deep dive into the constructor reveals parameters such as \lstinline|dataframe|, \lstinline|columns_to_process|, \lstinline|columns_to_deduplicate|, among others, which play a pivotal role in customizing the processor's behavior according to specific needs. For instance, the \lstinline|dataframe| parameter holds the data subjected to processing, while configurations such as \lstinline|custom_delimiter| and \lstinline|deduplication_threshold| offer granular control over text processing and data deduplication respectively.

\begin{lstlisting}[language=Python]
class BaseProcessor:
def init(self, dataframe: pd.DataFrame, columns_to_process: list[str] = None, columns_to_deduplicate: list[str] = None, ... ):
    self.dataframe = dataframe.copy()
...
\end{lstlisting}

\paragraph{Initialization (\_\_init\_\_)}
During the initialization phase, the \lstinline
|BaseProcessor| is configured by a series of parameters that distinctly influence its behavior across a multitude of operations. At the heart of this configuration lies parameters such as \lstinline
|dataframe|, \lstinline|columns_to_process|, and \lstinline|columns_to_deduplicate|, among others. For instance, the \lstinline|dataframe| parameter embodies the data destined for processing, acting as a receptacle that holds the information to be manipulated and refined.

\begin{lstlisting}[language=Python]
    def __init__(self, dataframe: pd.DataFrame, columns_to_process: list[str] = None, ...)
        # Implementation details
\end{lstlisting}

\paragraph{Data Validation and Preprocessing}
Preliminary steps within the \lstinline|BaseProcessor| pipeline include data validation and cleansing operations that are critical for maintaining the accuracy and quality of the analysis.

\textit{handle\_nan}
The \lstinline|handle_nan| method provides mechanisms for dealing with missing values, which can be pivotal for preserving the dataset's consistency, especially during deduplication phases.

\begin{lstlisting}[language=Python]
    def handle_nan(self, mode_type = 'deduplication')
        # Implementation details
\end{lstlisting}

\textit{validate\_dataframe}
The \lstinline|validate_dataframe| method is instrumental in ensuring the integrity of the data, validating the data frame's structure and type. This method is invoked during the initialization phase, ensuring that the data is compatible with the subsequent processing steps.

\begin{lstlisting}[language=Python]
    def validate_dataframe(self)
        # Implementation details
\end{lstlisting}

\paragraph{Custom Operations}
The \lstinline|BaseProcessor| extends its functionality by allowing the incorporation of bespoke operations, imparting versatility to the user's data processing needs.

\textit{apply\_custom\_operations}
The \lstinline|apply_custom_operations| method facilitates the execution of an array of user-defined transformations, enhancing the adaptability of the data processing workflow.

\begin{lstlisting}[language=Python]
    def apply_custom_operations(self, operations: list)
        # Implementation details
\end{lstlisting}

\textit{combine\_columns}
Merging multiple columns is accomplished via the \lstinline|combine_columns| method, a process that can be essential for feature engineering and dataset enrichment.

\begin{lstlisting}[language=Python]
    def combine_columns(self, columns_to_combine: list[str], new_col_name = 'target_col') -> None
        # Implementation details
\end{lstlisting}

\paragraph{Text Processing}
The \lstinline|BaseProcessor| module also houses sophisticated text manipulation utilities, providing the foundation for any text-centric analytical task.

\textit{split\_by\_delimiter}
Dividing text based on specified delimiters.
\begin{lstlisting}[language=Python]
    def split_by_delimiter(self, text: str) -> list[str]
        # Implementation details
\end{lstlisting}

\textit{tokenize\_string}
Tokenization, facilitated by the \lstinline|tokenize_string| method, parses strings into tokens, which are the basic methods for natural language processing (NLP) tasks.

\begin{lstlisting}[language=Python]
    def tokenize_string(self, text: list[str]) -> list[str]
        # Implementation details
\end{lstlisting}

\paragraph{Hyphen Handling}
Given the idiosyncrasies of text data, the \lstinline|BaseProcessor| adeptly manages hyphenated constructs to ensure uniformity in processed tokens.

\textit{handle\_hyphenated\_terms}
Dealing with terms that contain hyphens for standardized processing.
\begin{lstlisting}[language=Python]
    def handle_hyphenated_terms(self, tokens: list[str]) -> list[str]
        # Implementation details
\end{lstlisting}

\paragraph{Numerical/Special Character Handling}
A comprehensive text processing system must discern and manage numerical and special characters efficiently, as they can bear significant or negligible meaning depending on the context.

\textit{remove\_numbers}
The \lstinline|remove_numbers| method in \lstinline|BaseProcessor| offers a versatile approach to filtering out numerical values from textual data. This function can be tailored with different \lstinline|pattern_type| parameters to match specific patterns of numbers within the text.

\begin{lstlisting}[language=Python]
def remove_numbers(self, tokens: list[str], pattern_type: str = 'all') -> list[str]
    # Implementation details
\end{lstlisting}

\paragraph{Punctuation Handling}
Punctuation marks, while crucial for human readability, often add noise to text-based algorithms. Their management is therefore crucial in pre-processing.

\textit{remove\_punctuation}
The \lstinline|remove_punctuation| function is designed to strip text of unnecessary punctuation, enhancing the textual data's uniformity and reducing complexity for downstream processing.

\begin{lstlisting}[language=Python]
def remove_punctuation(self, text: str | list[str], punctuation_type = 'default') -> str | list[str]
    # Implementation details
\end{lstlisting}

\paragraph{Text Transformation}
The transformation of textual data is a critical pre-processing step to ensure uniformity and facilitate the application of NLP algorithms.

\textit{stem\_tokens}
The \lstinline|stem_tokens| method reduces words to their base or stem form, a crucial step in the normalization process that helps in decreasing the complexity of textual data by consolidating various forms of a word into a common base.

\begin{lstlisting}[language=Python]
def stem_tokens(self, tokens)
# Implementation details
\end{lstlisting}

\textit{to\_lowercase}
The conversion of all text entries to lowercase using the \lstinline|to_lowercase| function ensures case consistency throughout textual data, which is particularly significant when the case is not indicative of different meanings.

\begin{lstlisting}[language=Python]
def to_lowercase(self, text: str) -> str
    # Implementation details
\end{lstlisting}

\paragraph{Text Cleanup and Normalization}
Normalization is a process that not only cleans the text but also standardizes it, making it more amenable to analysis.

\textit{unicode\_normalize}
The \lstinline|unicode_normalize| function standardizes Unicode characters into their canonical form, thereby simplifying the encoding of text data.

\begin{lstlisting}[language=Python]
def unicode_normalize(self, text: str) -> str
    # Implementation details
\end{lstlisting}

\textit{remove\_non\_ascii}
By removing non-ASCII characters with the \lstinline|remove_non_ascii| method, the text is cleansed of characters that could lead to processing errors or inconsistencies.

\begin{lstlisting}[language=Python]
def remove_non_ascii(self, tokens)
    # Implementation details
\end{lstlisting}

\textit{remove\_angle\_brackets}
The \lstinline|remove_angle_brackets| method is employed to eliminate any text or markup that is enclosed within angle brackets, often used to denote tags or metadata that are not required for the main text analysis.

\begin{lstlisting}[language=Python]
def remove_angle_brackets(self, tokens)
    # Implementation details
\end{lstlisting}

\textit{strip\_whitespace}
Trimming extraneous whitespaces from the data via the \lstinline|strip_whitespace| function is an essential step in tidying the text and preparing it for further processing without the noise of unnecessary spaces.

\begin{lstlisting}[language=Python]
def strip_whitespace(self, tokens)
    # Implementation details
\end{lstlisting}

\textit{remove\_stopwords}
Stopwords are commonly occurring words in a language that carry minimal unique information. The \lstinline|remove_stopwords| method excises these to focus the analysis on the most meaningful words.

\begin{lstlisting}[language=Python]
def remove_stopwords(self, tokens)
    # Implementation details
\end{lstlisting}

\textit{final\_cleanup}
A comprehensive \lstinline|final_cleanup| method acts as a concluding step in the text pre-processing phase, ensuring that the text is in its most refined form before being passed on for analysis or machine learning processes.

\begin{lstlisting}[language=Python]
def final_cleanup(self, tokens: str) -> str
    # Implementation details
\end{lstlisting}

\paragraph{Execution}
The \lstinline|BaseProcessor| is architected to be versatile and extensible. The \lstinline|execute_processor| abstract method provides a template for the execution logic that must be implemented by derived classes, tailoring the processing pipeline to specific datasets and applications.

\textit{execute\_processor}
An abstract method that mandates implementation by subclasses, dictating the execution logic of the processing steps.
\begin{lstlisting}[language=Python]
def execute_processor(self) -> pd.DataFrame
    # This method must be implemented by subclasses to define the specific processing execution logic.
\end{lstlisting}

In summary, the \lstinline|BaseProcessor| is a robust and sophisticated module within the \lstinline|pyKCN|, capable of executing a diverse array of text pre-processing functions. Its comprehensive set of methods for text transformation and cleanup is foundational in preparing data for advanced NLP tasks, ensuring that downstream processing is conducted on data that is clean, normalized, and standardized.

\subsubsection{DeduplicationProcessor: Ensuring Data Uniqueness}
Data deduplication stands as an imperative process in the data preparation phase, particularly in tasks where the uniqueness of entries is a prerequisite for analytical accuracy. The \lstinline|DeduplicationProcessor|, inheriting the \lstinline|BaseProcessor|, epitomizes the commitment to data quality by removing redundancies through a customizable pipeline of pre-processing functions.

The \lstinline|DeduplicationProcessor| has been architected to refine textual data in preparation for deduplication. It undertakes a methodical sequence of normalization and transformation processes to eliminate inconsequential variations such as case sensitivity, punctuation discrepancies, and encoding differences. Subsequently, the harmonized data can be scrutinized for redundancies using string similarity assessments or advanced deduplication algorithms.

\paragraph{Initialization}
The instantiation of the \lstinline|DeduplicationProcessor| includes parameters tailored to identify and remove duplicates, taking into account a set threshold that determines the strictness of comparison for deduplication.

\begin{lstlisting}[language=Python]
    class DeduplicationProcessor(BaseProcessor):
        # ... constructor and DEDUPLICATION_STEPS omitted for brevity ...
    \end{lstlisting}

\paragraph{Similarity Assessment}
At the core of this processor is the capability to discern similarity between string entries, which is pivotal for identifying duplicates beyond exact matches.

\begin{lstlisting}[language=Python]
    @staticmethod
    def is_similar(string1: str, string2: str, threshold_percentage: int) -> bool:
        # Levenshtein Distance based similarity check (fuzzy matching)
    \end{lstlisting}

\paragraph{Deduplication Strategies}
The class defines two primary strategies for deduplication: exact match deduplication and fuzzy deduplication based on similarity thresholds, allowing for flexible data cleansing based on the use case.

\begin{lstlisting}[language=Python]
def deduplicate_based_on_similarity(self, ...):
    # Implementation details

def remove_duplicates(self, ...):
    # Implementation details
\end{lstlisting}

\paragraph{Execution Pipeline}
The execution method is where the deduplication process is orchestrated, utilizing the defined pre-processing steps to prepare data before applying the deduplication logic.

\begin{lstlisting}[language=Python]
def execute_processor(self) -> pd.DataFrame:
    # Implementation details
\end{lstlisting}

\paragraph{Deduplication Pipeline}
Each step in the deduplication process is concisely described, and the operations are applied in sequence to the specified columns of the DataFrame. The methodological removal of duplicates ensures that the dataset retains only the essential information, minimizing data redundancy.

The operational sequence of the \lstinline|DeduplicationProcessor| is composed of a series of methodical steps aimed at cleansing and normalizing text data. These steps include:

\begin{itemize}
    \item Removal of content within angle brackets that often signifies irrelevant metadata or formatting.
    \item Conversion of all text to a uniform lowercase format to eliminate case-based discrepancies.
    \item Application of Unicode normalization to resolve encoding variances.
    \item Purging of non-ASCII characters to maintain a standard character set.
    \item Exclusion of punctuation marks to reduce textual noise and facilitate comparison.
    \item Eradication of numerical figures where they are deemed non-essential for the analysis context.
\end{itemize}

Each step in this deduplication pipeline is paramount in its own right, coalescing into a powerful collective that refines text into a form optimized for duplication assessment.

\paragraph{Usage and Examples}
The \textit{DeduplicationProcessor} can be instantiated and employed in various contexts, ranging from simple use cases to complex data preparation pipelines. The following examples provide a glimpse into its integration within a data processing workflow.

\begin{lstlisting}[language=Python]
# Example usage of DeduplicationProcessor
deduplication_processor = DeduplicationProcessor(dataframe, columns_to_process=['column_1', 'column_2'])
processed_df = deduplication_processor.execute_processor()
\end{lstlisting}

\paragraph{Algorithmic Implementation}
At the code level, the \lstinline|DeduplicationProcessor| is a Python class that inherits from the \lstinline|BaseProcessor|. It is initialized with parameters that define the scope of deduplication, the threshold for fuzzy matches, and the steps to be executed in the deduplication pipeline. The class methods encompass functionality for string similarity checks, dataframe deduplication based on exact and fuzzy matches, and the primary \lstinline|execute_processor| method, which encapsulates the full deduplication logic.

\paragraph{Summary}
In summation, the \lstinline|DeduplicationProcessor| stands as a testament to the software's capabilities in ensuring data uniqueness and integrity. It highlights the intricate blend of text processing techniques, from normalization to fuzzy matching, to address the ubiquitous challenge of data redundancy. The inclusion of this module within the \lstinline|pyKCN| suite reinforces the tool's comprehensive approach to data quality management.

\subsubsection{TextProcessor: Specialized Textual Data Processing}
The \lstinline|TextProcessor| class, a derivative of the \lstinline|BaseProcessor|, embodies a more focused approach towards textual data. It accentuates the foundational pre-processing capabilities with specialized pipelines that cater to text-specific operations such as stemming, normalization, and removal of unwanted characters. 

\paragraph{Primary and Stemming Pipelines}
The \lstinline|TextProcessor| is equipped with two default pipelines: the \lstinline|DEFAULT_PRIMARY_PIPELINE| and the \lstinline|DEFAULT_STEMMING_PIPELINE|. These structured sequences of operations are tailored to refine textual data methodically. The primary pipeline concentrates on pre-cleanup and preparation tasks, while the stemming pipeline handles the reduction of words to their base forms and further cleanup.

\begin{lstlisting}[language=Python]
    class TextProcessor(BaseProcessor):
        # Class definition and pipelines as provided.
    \end{lstlisting}

\paragraph{Constructor Details}
Upon instantiation, the \lstinline|TextProcessor| initializes with an extensive set of parameters, offering extensive control over the processing procedures. This includes specifying columns to process, setting a deduplication threshold, and configuring pipeline-related settings, among others.

\begin{lstlisting}[language=Python]
    def __init__(self, dataframe: pd.DataFrame, columns_to_process: list[str], ...):
        super().__init__(dataframe, columns_to_process, ...)
        # Additional initialization details.
    \end{lstlisting}

\paragraph{Execution Logic}
The execution methods, \lstinline|execute_processor|, \lstinline|execute_primary_pipeline|, and \lstinline|execute_stemming_pipeline|, articulate the logic for the sequential processing of data through the defined pipelines.

\begin{lstlisting}[language=Python]
    def execute_processor(self, ...):
        # Execution logic for processing pipelines.
    \end{lstlisting}

\paragraph{Predefined Pipelines}
The \lstinline|TextProcessor| class comes with two predefined pipelines that establish a sequence of operations to be applied to the textual data: 

the \lstinline|DEFAULT_PRIMARY_PIPELINE| and the \lstinline|DEFAULT_STEMMING_PIPELINE|. These pipelines are collections of processing steps, each defined as a dictionary object that specifies the operation to be executed and its associated arguments.

\textit{DEFAULT\_PRIMARY\_PIPELINE}
The \lstinline|DEFAULT_PRIMARY_PIPELINE| is a sequence designed for the initial cleanup and normalization of text. It includes steps such as the removal of markup enclosed within angle brackets, case normalization, Unicode normalization, tokenization, non-ASCII character removal, and more. Each step is meticulously described and assigned a specific operation function within the \lstinline|TextProcessor|.

\begin{lstlisting}[language=Python]
DEFAULT_PRIMARY_PIPELINE = {
    'default': [
        # ... Pipeline steps as described.
    ]
}
\end{lstlisting}

\begin{itemize}
    \item Removing text within angle brackets.
    \item Splitting text based on delimiters for tokenization.
    \item Converting text to lowercase to ensure case consistency.
    \item Normalizing Unicode characters to their canonical form.
    \item Tokenizing strings into individual terms.
    \item Removing non-ASCII characters for standardization.
    \item Handling hyphenated terms for consistent processing.
    \item Eliminating common stopwords.
    \item Stripping extraneous whitespace from the text.
    \item Discarding numerical values based on specified patterns.
    \item Purging punctuation, configurable by type.
\end{itemize}

These steps are crucial for preparing the text for advanced processing tasks, ensuring that the data is consistent and standardized.

\textit{DEFAULT\_STEMMING\_PIPELINE}

After the primary processing, the \lstinline|DEFAULT_STEMMING_PIPELINE| targets the stemming of words and additional cleanup. Stemming reduces words to their root form, significantly reducing the complexity of language by consolidating variations of a word. This pipeline also includes rejoining terms and filtering by length, which removes extraneous or overly short terms that may not contribute meaningful information to the analysis.

\begin{lstlisting}[language=Python]
DEFAULT_STEMMING_PIPELINE = {
    'original_data': [
        # ... Steps for processing original data.
    ],
    'stemmed_data': [
        # ... Steps for processing stemmed data.
    ]
}
\end{lstlisting}

\begin{itemize}
    \item Rejoining terms post tokenization for original data.
    \item Filtering terms based on their length.
    \item Cleaning the text of any remaining unnecessary characters in original data.
    \item Applying stemming to tokenize terms.
    \item Rejoining stemmed terms into a coherent text.
    \item Further filtering by length for stemmed data.
    \item Final cleanup of stemmed terms.
\end{itemize}

\paragraph{Integration of Pipelines}
During the initialization of the \lstinline|TextProcessor|, these pipelines are set as defaults, but they can be replaced or augmented by user-defined pipelines, offering flexibility and customizability.

\begin{lstlisting}[language=Python]
def __init__(self, ...):
    self.PRIMARY_PIPELINE = primary_pipeline or self.DEFAULT_PRIMARY_PIPELINE
    self.STEMMING_PIPELINE = stemming_pipeline or self.DEFAULT_STEMMING_PIPELINE
    # Additional initialization logic...
\end{lstlisting}

The primary and stemming pipelines are key components of the \lstinline|TextProcessor|, allowing for a structured and sequential approach to text processing. The default configurations of these pipelines reflect best practices in text preprocessing and are optimized for a general use case, ensuring that the \lstinline|TextProcessor| is versatile and ready to handle diverse textual datasets right out of the box.

\paragraph{Caching Mechanism}
An integral feature of the \lstinline|TextProcessor| is its caching mechanism, which is meticulously designed to store intermediate results, reducing computational overhead in repeated operations.

\begin{lstlisting}[language=Python]
def save_cached_data(self, pipeline_type: str):
    # Code to save DataFrame to cache.
\end{lstlisting}

The caching mechanism's load operations are equally optimized to retrieve stored data, ensuring efficient continuity in data processing.

\begin{lstlisting}[language=Python]
def load_cached_data(self, pipeline_type: str):
    # Code to load DataFrame from cache.
\end{lstlisting}

\paragraph{Utility Methods}
The class also includes utility methods like \lstinline|get_cache_file_path| and \lstinline|_handle_parquet_format|, which support the caching system and facilitate the management of cache files.

\begin{lstlisting}[language=Python]
def get_cache_file_path(self, pipeline_type: str):
    # Code to get the file path for the cache.
\end{lstlisting}

These utility methods ensure that the \lstinline|TextProcessor| operates seamlessly across different stages of the text processing lifecycle.

\paragraph{Data Retrieval}
Post-processing, the \lstinline|TextProcessor| provides methods to access the processed data, thereby encapsulating the entire data processing workflow within a coherent interface.

\begin{lstlisting}[language=Python]
def get_primary_pipeline_data(self):
    # Code to retrieve data processed by the primary pipeline.
\end{lstlisting}

\paragraph{Summary}
The \lstinline|TextProcessor| class is a testament to advanced textual data handling, integrating a comprehensive set of preprocessing functionalities with an efficient caching system. This orchestration of operations not only amplifies the processing capabilities but also ensures that the transformed data adheres to the highest standards of quality and readiness for downstream applications such as machine learning modeling and data analysis.

\section{Downstream Tasks}
The processed data output by the \lstinline|TextProcessor| module serves as a cornerstone for several downstream tasks. These tasks leverage the sanitized and structured data to unearth patterns, associations, and insights that are pivotal for knowledge discovery within vast repositories of textual information, such as scientific articles. We focus on tasks that reveal the intricacies of keyword interrelationships, deduce association rules, offer general statistics, and explore innovative applications of GPT-4.

\subsection{Downstream Tasks}

\subsubsection{Network Metrics/Parameters}

Weighted network metrics evaluate keyword connections on different levels of significance, relevance, and hierarchies. Given a keyword network consists of entities interconnected with non-zero probability, the weighted network metrics can help the KCN quantify the evolution in a knowledge domain. Previous research conducted by Barrat et al. and  Duvuru et al. has presented weighted network metrics that can reflect emerging trends and knowledge structures. Radhakrishnan et al. also developed novel weighted network metrics and validated the effectiveness in nano-related literature research. Accordingly, several key network metrics are summarized as follows:

The level of degree is a fundamental measurement of node centrality, which indicates its importance. Considering $k_i$ as the level of degree for a node $i$, and the group of links $a_{ij}$, whose elements take the value 1 if a link connects node $i$ to the node $j$ and 0 otherwise, the level of degree can be denoted as:
\begin{equation}
k_i = \sum_{j \in N_i} a_{ij}
\end{equation}

While the links hold different significance, the strength of node is more subjective performance evaluation for the node centrality. Assuming only symmetric weights exist among the nodes, a matrix that keeps the values of weighted links $w_{ij}$ should fulfill that $w_{ij} = w_{ji}$, which indicates the weight of the link between node $i$ and $j$ is the same regardless of the direction. The strength $s_i$ can be further denoted as:
\begin{equation}
s_i = \sum_{j \in N_i} w_{ij}
\end{equation}

Either strong or weak connections may exist among nodes with different levels of degrees. The average weight of endpoint degrees was introduced to measure the co-occurrence of connections between pairs of nodes as their degrees are changing. Considering $k_i$ and $k_j$ are the degrees of node $i$ and $j$, respectively, the product of $k_i$ and $k_j$ is defined as the average weight of endpoint degrees:
\begin{equation}
    <w_{ij}> = k_i k_j
\end{equation}

If a node and its neighbors share a similar level of degree, an enrichment tendency might exist and can be reflected by an index of the average weighted nearest neighbor's degrees. Given a set of notes $N_i$, the index  is defined as:
\begin{equation}
    k^w_{nn,i} = \frac{1}{s_i} \sum_{j \in N_i} w_{ij}k_j
\end{equation}
Where the index $k^w_{nn,j}$ can also be considered as the affinity measurement that indicates the tendency of nodes to connect with the nearest nodes that contain similar degree levels. The network is determined as assortative if this measurement is proportional to degree or vice versa. 

 Local cohesiveness of groups of nodes can reflect how well a particular node is connected to its neighbors and it can be measured by a weighted clustering coefficient. This measurement weighs a single node's surrounding structure on the intensity of its interaction with the local triplets. Considering a node $i$, along with its connected neighbors $N_i$, degree $k_i$ and strength $s_i$, the weighted clustering coefficient $C^w_{i}$ can be denoted as:
 
\begin{equation}
C_i^w = \frac{1}{s_i(k_i - 1)} \sum_{j \in N_i,\\ h \in N_i} \left( \frac{w_{ij} + w_{ih}}{2} \right) a_{jh}
\end{equation}

\subsection{Pain Research}
We employed this methodology in the field of pain research, which has been gaining global significance due to the increasing number of individuals experiencing pain-related issues. In response to this public health challenge, various interdisciplinary research areas have converged to address pain-related concerns. This convergence has led to a rapid increase in the number of studies and has placed greater demands on researchers. Therefore, this study conducts a comprehensive review and analysis of a substantial body of pain-related literature using the keyword co-occurrence network (KCN) methodology.

For this investigation, we extracted and examined keywords from 264,560 pain-related research articles indexed in IEEE, PubMed, Engineering Village, and Web of Science, published between 2002 and 2021. We created four distinct networks representing four different time windows and applied network metrics to identify frequently used keywords related to pain, reveal patterns of association among these pain-related keywords, and examine the research trends within the domains of sensors/methods, biomedical research, and treatment approaches in the context of pain-related studies.
\cite{ozek2023analysis} 

\subsection{Asset Life Cycle Management Research}
We then applied this approach to explore the progression of Industry 4.0 technology applications within the realm of sustainable asset life cycle management (ALCM). The increasing potential of Industry 4.0 technologies to promote sustainable manufacturing has generated a growing emphasis on ALCM in recent years.

The foundation of this study lies in the analysis of keywords extracted from 3,896 scientific articles related to ALCM. These articles were published in Web of Science, IEEE Xplore, and Engineering Village during the period from 2002 to 2021. We constructed KCNs based on these keywords and conducted an in-depth exploration of network characteristics to reveal valuable knowledge patterns, components, structure, and research trends.\cite{weerasekara2022trends}

\subsection{AI-assisted Vehicle Maintenance} 
The increasing complexity of a vehicle's digital architecture has created new opportunities to revolutionize the maintenance paradigm. The Artificial Intelligence (AI) assisted maintenance system is a promising solution to enhance efficiency and reduce costs. 

The KCN methodology is applied to systematically analyze the keywords extracted from 3153 peer-reviewed papers published between 2011 and 2022. The network metrics and trend analysis uncovered important knowledge components and structure of the research field covering AI applications for vehicle maintenance. The emerging and declining research trends in AI models and vehicle maintenance application scenarios were identified through trend visualizations.  \cite{li2023study}

% \section{Example, Breast Cancer}

% \textbf{Breast Cancer Example}

% \section{Discussion - All} \label{sec:discussion}

% \section{Future Work - All} \label{sec:future}

% \begin{leftbar}
% \dots \dots
% \end{leftbar}

\section*{Acknowledgments}

\begin{leftbar}
Our teams have been published in the following papers based on this work: \cite{ozek2023analysis, weerasekara2022trends, li2023study, li2024navigating}
\end{leftbar}

\small
\bibliographystyle{ieee_fullname}
\bibliography{pyKCN
}

\begin{thebibliography}{1}\itemsep=-1pt

\bibitem{li2023study}
Wei Li, Guoyan Li, and Sagar Kamarthi.
\newblock The study of trends in ai applications for vehicle maintenance
  through keyword co-occurrence network analysis.
\newblock {\em International Journal of Prognostics and Health Management},
  14(2), 2023.

\bibitem{li2024navigating}
Wei Li, Haozhou Zhou, Zhenyuan Lu, and Sagar Kamarthi.
\newblock Navigating the evolution of digital twins research through keyword
  co-occurence network analysis.
\newblock {\em Sensors}, 24(4):1202, 2024.

\bibitem{ozek2023analysis}
Burcu Ozek, Zhenyuan Lu, Fatemeh Pouromran, Srinivasan Radhakrishnan, and Sagar
  Kamarthi.
\newblock Analysis of pain research literature through keyword co-occurrence
  networks.
\newblock {\em PLOS Digital Health}, 2(9):e0000331, 2023.

\bibitem{pdfai2023}
PDF.ai.
\newblock {pdf.ai}: Chat with any pdf document.
\newblock {\em https://www.pdf.ai}, 2023.
\newblock Accessed: 2023-11-01.

\bibitem{radhakrishnan2017novel}
Srinivasan Radhakrishnan, Serkan Erbis, Jacqueline~A Isaacs, and Sagar
  Kamarthi.
\newblock Novel keyword co-occurrence network-based methods to foster
  systematic reviews of scientific literature.
\newblock {\em PloS one}, 12(3):e0172778, 2017.

\bibitem{van2010software}
Nees Van~Eck and Ludo Waltman.
\newblock Software survey: Vosviewer, a computer program for bibliometric
  mapping.
\newblock {\em scientometrics}, 84(2):523--538, 2010.

\bibitem{Ammar2018ConstructionOT}
Chandra~Bhagavatula Waleed~Ammar, Dirk~Groeneveld.
\newblock Construction of the literature graph in semantic scholar.
\newblock {\em North American Chapter of the Association for Computational
  Linguistics}, 2018.

\bibitem{weerasekara2022trends}
Sachini Weerasekara, Zhenyuan Lu, Burcu Ozek, Jacqueline Isaacs, and Sagar
  Kamarthi.
\newblock Trends in adopting industry 4.0 for asset life cycle management for
  sustainability: a keyword co-occurrence network review and analysis.
\newblock {\em Sustainability}, 14(19):12233, 2022.

\end{thebibliography}
%%%%%%%%%%%%%%%%%%%%%%%%%%%%%%%%%%%%%%%%%%%%%%%%%%%%%%%%%%%%

\end{document}